\documentclass[letterpaper,twocolumn,10pt]{article}
\usepackage{usenix,epsfig,endnotes}
\usepackage{appendix}
\usepackage{cite}
\usepackage{amsmath,amssymb,amsfonts}
\usepackage{algorithmic}
\usepackage{graphicx}
\usepackage{textcomp}
\usepackage{xcolor}
\usepackage{times}
\usepackage{helvet}
\usepackage[ruled,vlined,linesnumbered]{algorithm2e}
\usepackage{mathtools}
\usepackage{tabularx}
\usepackage{booktabs}
\usepackage[english]{babel}
\usepackage{listings}
\usepackage{array,multirow}
\usepackage{float}
\usepackage{pifont}
\usepackage{verbatim} 
\usepackage{enumitem}
\usepackage{rotating}
\usepackage{adjustbox}
\usepackage[normalem]{ulem}
\usepackage{amsmath}
\usepackage{makecell}
\usepackage{tikz}
\usepackage{subcaption}
\usepackage{amsmath}

\usetikzlibrary{fit,arrows,calc,positioning}
\usetikzlibrary{shapes,arrows}

\newcommand{\methodname}{Autosploit\xspace}

\usepackage[utf8]{inputenc}
\usepackage[english]{babel}
\usepackage{amsthm}
\usepackage{verbatim}
\usepackage{authblk}
\newcommand*{\affaddr}[1]{#1} 
\newcommand*{\affmark}[1][*]{\textsuperscript{#1}}

\theoremstyle{definition}
\newtheorem{definition}{Definition}[]
\theoremstyle{remark}
\newtheorem{example}{Example}[]
\theoremstyle{definition}

\def\BibTeX{{\rm B\kern-.05em{\sc i\kern-.025em b}\kern-.08em
    T\kern-.1667em\lower.7ex\hbox{E}\kern-.125emX}}
\begin{document}

\date{}

\title{\Large \bf \methodname: A Fully Automated Framework for Evaluating \\ the Exploitability of Security Vulnerabilities}

\author{%
Noam Moscovich\affmark[1], Ron Bitton\affmark[1], Yakov Mallah\affmark[1], Masaki Inokuchi\affmark[2], Tomohiko Yagyu\affmark[2], Yuval Elovici\affmark[1] and Asaf Shabtai\affmark[1]\\
\affaddr{\affmark[1]Dept. of Software and Information Systems Engineering, Ben-Gurion University of the Negev }\\
\affaddr{\affmark[2]NEC Corporation}\\
}

\maketitle

\begin{abstract}
The existence of a security vulnerability in a system does not necessarily mean that it can be exploited.
In this research, we introduce \methodname \textemdash an automated framework for evaluating the exploitability of vulnerabilities.
Given a vulnerable environment and relevant exploits, \methodname will automatically test the exploits on different configurations of the environment in order to identify the specific properties necessary for successful exploitation of the existing vulnerabilities.
Since testing all possible system configurations is infeasible, we introduce an efficient approach for testing and searching through all possible configurations of the environment. 
The efficient testing process implemented by \methodname is based on two algorithms: \textit{generalized binary splitting} and \textit{Barinel}, which are used for noiseless and noisy environments respectively.
We implemented the proposed framework and evaluated it using real vulnerabilities.
The results show that \methodname is able to automatically identify the system properties that affect the ability to exploit a vulnerability in both noiseless and noisy environments.
These important results can be utilized for more accurate and effective risk assessment.  
\end{abstract}

\section{\label{sec:intro}Introduction}
Risk assessment is an activity essential for improving the security of an enterprise network~\cite{landoll2005security}. 
A traditional cyber security risk assessment procedure begins with identifying system assets and enumerating the threats to which those assets are exposed. 
Next, \textit{vulnerability assessment} tools are used to reveal the existence of security vulnerabilities in system components. 
This process is usually performed using \textit{vulnerability scanners} (such as Nessus~\cite{Nessuscite} or OpenVAS~\cite{developers2012open}) aimed at identifying components or services that are vulnerable to known attacks~\cite{nilsson2006vulnerability}.
Based on the security vulnerabilities identified and the enterprise's network topology, the attack surface can be generated~\cite{Albanese2014ManipulatingTA}. 
The attack surface represents the possible attack paths an attacker can take to compromise an asset. 
Finally, given the attack surface, an optimal mitigation strategy (such as hardening the system) can be implemented to minimize the overall risk to the system.

The correctness and completeness of the attack surface is highly dependent on the inputs from the vulnerability assessment phase.
The use of vulnerability scanners to identify the presence of security vulnerabilities in a target system can be performed automatically and continuously, and is relatively safe (cannot harm the target system), since these tools do not actually perform any exploitation.

However, vulnerability assessment that is solely based on vulnerability scanners has one main limitation: it cannot guarantee that the vulnerabilities identified can be exploited.
This is because the exploitability of a security vulnerability depends on the specific properties (i.e., configuration) of the vulnerable system, which is not considered during the vulnerability scanning process.
Examples of such properties are the libraries installed, the services running (which are not bound to the network), and the implementation of safeguards, such as stack protectors.

Considering vulnerabilities that cannot be exploited in the attack surface may enumerate attack paths that cannot be executed by an attacker, which results in an incorrect allocation of resources for countermeasures.
Therefore, to conduct an effective risk assessment process, it is essential to evaluate the exploitability of known vulnerabilities for the specific system configuration. 

The most common approach for evaluating the exploitability is penetration testing (PT), a goal-oriented simulation of a cyber attack on a specific network, system, or software application.
PT aims at evaluating and demonstrating the exploitability of known vulnerabilities, as well as identifying unknown vulnerabilities.
However, PT is a time and resource-consuming task, usually performed manually by security experts (using tools, such as Kali Linux~\cite{denis2016penetration}), and therefore, in practice, it cannot be performed continuously.

A complementary method for evaluating exploitability is the Common Vulnerability Scoring System (\textit{CVSS}), an open and widely adopted framework for quantifying the severity of software vulnerabilities \cite{mell2006common}.
Given a software vulnerability, the CVSS provides a numerical score indicating the severity, impact, and exploitability of the vulnerability.
However, this score is calculated for each vulnerability regardless of the specific system configuration.
It should be mentioned that the CVSS provides a mathematical framework that can be used to adjust the severity, impact, and exploitability metrics based on the specific system configuration.
In particular, the environmental metrics enable security analysts to customize the CVSS score based on specific characteristics of a user's environment (such as characteristics that affect exploitability).
However, the CVSS does not determine the characteristics that are necessary for exploiting the vulnerability or provide any guidelines for/means of measuring those characteristics given a specific system configuration, thus leaving the security analyst with no practical solution.
To the best of our knowledge, an automated tool that can be used by security analysts to measure those characteristics, given a specific system configuration, has yet to be proposed.

In this paper, we introduce \methodname \textemdash~an automated framework for evaluating the exploitability of security vulnerabilities.
Given a vulnerable environment and relevant exploit, \methodname will automatically test the exploit in different system configurations in order to identify the specific properties (e.g., installed packages, running services) that are necessary to successfully exploit the vulnerability.
This is done using a dedicated simulator that automates the process of changing the system configuration, exploiting the system, and collecting evidence with respect to the success of the exploitation.  
Since modern environments consist of multiple configurable parameters, a straightforward approach in which all possible parameters are tested individually is very time-consuming and therefore impractical.
The proposed method takes an efficient approach for testing and searching through the system configurations. 
The approach is based on two algorithms: generalized binary splitting~\cite{ding-zhudu1993} and Barinel~\cite{abreu2009spectrum}, which are used for noiseless and noisy environments respectively.
We evaluated the proposed framework by testing the exploitability of real vulnerabilities on different configurations of the Linux-based Metasploitable 2 machine~\cite{moore2012metasploitable}. 
Our results show that \methodname is able to identify the exact system configurations that affect the exploitability of the system in both noiseless and noisy environments. 

\smallskip
\noindent We summarize our paper’s contributions as follows:
\begin{enumerate}
    \item We present a fully automated framework that can be used to evaluate the exploitability of security vulnerabilities given a specific environment, thereby ensuring a more accurate risk management process.
    \item The proposed framework can identify the specific properties that are necessary for successful exploitation of a given vulnerability. 
   By knowing those properties, security analysts can easily identify the hardening techniques required to reduce the attack surface.
   \item We suggest two algorithms that can be used to efficiently search through various system configurations, and given a security vulnerability, are able to identify the exact system configurations that affect the exploitability of the system in both noiseless and noisy environments.
    The two algorithms are adapted from the domains of group testing and software code diagnostics.
    \item We develop a dedicated simulator that automates the process of changing the system configuration, exploiting the system, and collecting evidence with respect to the success of the exploitation.

\end{enumerate}

The remainder of the paper is organized as follows:
In Section~\ref{sec:background}, we provide a brief introduction to group testing, and the generalized binary splitting and Barinel algorithms.
Then, we formally define the notations and terms for the problem in Section~\ref{sec:problem} and determine the set of system configurations that can potentially affect exploitability in Section~\ref{sec:system_har}.
In Section~\ref{sec:prop_method}, we present the proposed framework, providing a description of the  simulator developed and the algorithms used to efficiently search through the possible system configurations.
The evaluation and results of the proposed method are presented in Section~\ref{sec:evaluation}.
Finally, in Section~\ref{sec:future_work}, we conclude our work and suggest new research directions.

\section{\label{sec:background}Background}
In this section, we provide a brief introduction to group testing and describe the generalized binary splitting and Barinel algorithms.

\subsection{\label{sec:group_testing_overview}Group Testing}
Group testing is the process of identifying a subset of defective items among a large set of items.
Testing each item in the group individually to determine whether it is defective or not is inefficient -- $O(n)$ for noiseless (deterministic) environments and $O(\lambda^{-1} n)$ for noisy environments (in which the same test may produce different results), where $\lambda$ denotes the probability for a test to produce incorrect results and $n$ is the number of items.
In order to expedite the process of identifying defective items, group testing algorithms test groups of items simultaneously.
The underlying assumption in group testing is that the number of defective items is relatively small (even asymptotically small) compared to the total number of items.
Hence, it is possible to reduce the number of operations by testing subgroups of items.
The intuition for testing subgroups is that if a test is negative, then all items in the subgroup are considered to be flawless (or flawless with a high probability in noisy environments), and therefore can be eliminated from future examinations, thus accelerating the search.

Group testing problems can be independently classified into three categories: \textit{(i)} probabilistic or combinatorial, \textit{(ii)} adaptive or non-adaptive, and \textit{(iii)} noisy or noiseless.
The first category refers to the a priori knowledge available on the distribution of defective items: no knowledge (combinatorial) or some probability distribution (probabilistic).
Non-adaptive group testing determines all tests in advance, while adaptive group testing uses previous test results to determine the next test. 
Finally, noiseless group testing assumes that the result of a group test is accurate.
On the other hand, noisy group testing assumes that the result of a group test can be incorrect with some probability.

Binary splitting is a well-known procedure to identify a \textit{single} defective item in a contaminated set of items.
The algorithm partitions a set of $n$ items into two disjoint groups, such that neither group's size exceeds $2^{\lceil log n \rceil -1}$.
Next, the algorithm tests a group in order to determine which of the groups is contaminated (i.e., includes defective items).
Then, it recursively applies binary splitting on the contaminated group.
The number of tests required for identifying a \textit{single} defective item using binary splitting is $O(\lceil \log{n} \rceil)$.

In this research, we adapt the generalized binary splitting~\cite{ding-zhudu1993,hwang1972method}
algorithm to enumerate the environmental conditions (configurations) that affect the exploitability of a system in a noiseless environment.

\subsection{\label{sec:barinel}Software Code Diagnosis}
A significant part of the software debugging process is the identification of faults and their location in the program code.
Although this process is crucial for ``healthy'' code, it is very expensive~\cite{tassey2002economic}.
These days, the scale of software code, combined with its complexity, makes manual debugging almost impossible.
Consequently, there is a need for an automated debugging tool.

Automatic fault localization techniques are a collection of procedures for automatically identifying the faulty components that cause a bug. 
Model-based diagnosis (MBD) is a known approach for automated diagnosis that has also been proposed for software diagnosis~\cite{stumptner96aModelBased}. 
In MBD, a model of the system is needed, along with observations of the system's behavior.
These observations are checked against the given model, and inference algorithms are used to produce \textit{diagnoses}, which are possible assumptions about which components are faulty but consistent with the given model and observations.
However, software systems can rarely be modeled accurately, and thus, directly applying MBD to software diagnosis is unscalable~\cite{wotawa2002creatingmodelformbd}. 

Software fault localization (SFL) is a general approach for fault diagnosis. 
In the context of software diagnosis, SFL operates by collecting the \textit{traces} (i.e., set of components involved in the execution) of the tests that were run.  
SFL methods for software diagnosis suggest diagnoses by considering the correlation between the passing and failing of tests and their traces. 

Barinel~\cite{abreu2009spectrum}
is an algorithm that falls between MBD and SFL that is specifically designed for software diagnosis. In this research, we adopt the Barinel algorithm to enumerate the environmental conditions that affect a system's exploitability.  
Barinel's features allow us to identify the environmental conditions that affect the exploitability of a system in a noisy environment.

\section{\label{sec:problem}Problem Definition}
This section formally defines the notations and terms for the problem of searching a subset of \textit{environmental conditions} that are necessary for the exploitation, using exploit $x$ of a vulnerability $v$.

\begin{definition}
Environmental conditions are specific parameters of a system environment that can potentially affect the exploitability of a vulnerability $v$ with exploit $x$. 
The set of environmental conditions is denoted by $C = \{c_1,...,c_n\}$.
 \end{definition}

\begin{definition}
The set of environmental conditions that are \textit{necessary} for the exploitation of $v$ by $x$ is represented as $R = \{r_1,...,r_n\}\in \{0,1\}^n$,
where $r_j=1$ if environmental condition $c_j$ is \textit{necessary} for the exploitation and $r_j=0$ otherwise.
\end{definition}

\begin{example}
The MS-RPC functionality in Samba version 3.0.0 allows a remote attacker to execute arbitrary commands via shell meta-characters (CVE-2007-2447). 
The exploitation of this vulnerability can be done by utilizing $ruby$ or $perl$ interpreters (installed in the victim environment) to create a reverse socket for initiating an interactive shell from the victim environment to the attacker environment.
Following that, $ruby$ and $perl$ are specific characteristics of a user’s environment that are necessary for the successful exploitation of the vulnerability.
\end{example}

\begin{definition}
A tested environment is represented by  $\boldsymbol{E}=\{e_1,...,e_n\} \in \{0,1\}^n$, where $e_j=1$, if the environmental condition $c_j$ is enabled in the environment and $e_j=0$ otherwise.
\end{definition}

\begin{figure}
    \centering
    \includegraphics[width=0.45\textwidth]{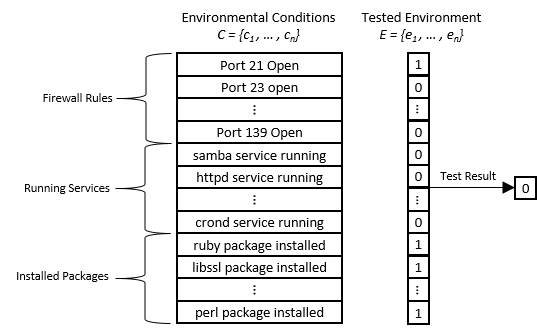}
    \caption{Illustrating the problem definition.}
    \label{fig:env_state}
\end{figure}

\begin{definition}
\label{def:test}
A \textit{test}, represented by the tuple $\langle E,result\rangle$, is an attempt (trail) for applying $x$ to exploit $v$ given an environment $E$.
We distinguish between two types of testing: testing in a noiseless (deterministic) environment and testing in a noisy environment.
When testing in a noiseless environment, the result of a test is positive (equal to $1$) if $\exists j: r_j=e_j=1$ and negative (equal to $0$) otherwise.
When testing in a noisy environment, the result of a test may be negative (equal to $0$) with some probability $\epsilon$ even when $\exists j : r_j=e_j=1$.
\end{definition}

\begin{example}
Figure~\ref{fig:env_state} illustrates a specific test.
The set of environmental conditions considered in the test are firewall rules (e.g., port $21$ is open), services running (e.g., $httpd$ service is running), and packages installed (e.g., $ruby$ package installed). 
The tested environment outlines the specific setup of the environment, subject to the environmental conditions.
For example, $e_1=1$ and $e_2=0$ indicate that the firewall within the tested environment blocks communication via port $21$ and allows communication via port $23$.
The result of the test ($result = 0$) indicates that the vulnerability $v$, to which the tested environment is exposed, can be exploited by $x$.    
\end{example}

\noindent\textbf{Objective.}
Find $\boldsymbol{R}$ with a minimal number of tests.

\section{\label{sec:system_har}Determining the Environmental Conditions}

Modern environments consist of multiple configurable parameters that may influence the security of the system (i.e., the ability to exploit existing vulnerabilities).
Many of these parameters (i.e., environmental conditions) are  used for operating system hardening.
Thus, in order identify them, we reviewed existing standards and guidelines for system hardening (e.g., \cite{scarfone2008guide,turnbull2006hardening,hassell2006hardening,mp2016enhancing}).
The list of environmental conditions identified is summarized in Appendix~\ref{sec:appendix}.
The environmental conditions are classified into five groups as follows: 

\begin{description}[style=unboxed,leftmargin=0cm]
\item[Access control:] 
The exploitability often depends on access control misconfigurations.
For instance, many exploits utilize account misconfigurations, such as the use of weak passwords,  guest users, or accounts with empty passwords, to compromise the system.
File permissions are another example, since many exploits utilize unauthorized data access to compromise a system.
Removing these user accounts, using a strong password policy, and correctly managing file permissions (i.e., the principle of least privileges) reduces a system's attack surface.

\item[Connectivity:]
The ability to exploit a vulnerability remotely (i.e., network attack vector) is heavily dependent on the connectivity settings. 
For example, blocking incoming communication to a vulnerable application may prevent the ability to exploit it; blocking the outgoing communication from a target environment to the attacker machine can prevent remote shell attacks.
In addition, many attacks utilize the ICMP protocol to compromise a system (e.g., flooding attacks).
For these reasons, restricting communication (e.g., using a firewall) can reduce the attack surface and thus, directly affects the exploitability of a system.

\item[Services:]
Exploitation often depends on the services running.
For instance, the default configurations of $ftp$ and $rsh$ services transfer usernames and passwords in plain-text, which can be captured by an attacker.
Other examples are the $telnet$ and $nis$ services, which enable remote login without authentication.
Removing/disabling these services reduces the attack surface and thus, directly affects the exploitability of a system.

\item[Safeguards:] Modern operating systems are equipped with many security features, which can be used to prevent the exploitation of vulnerable applications.
For example, data execution prevention (DEP~\cite{team2003pax}) prevents code execution from data segments; address space layout randomization (ASLR~\cite{team2003address}) makes it more difficult for an attacker to predict target addresses of a subroutine that is already present in the executable memory of a process.
Stack protectors detect buffer overflows on stack variables (by using the canary value~\cite{cowan1999protecting}).
Enabling these security features makes exploitation more difficult, thus directly affecting a system's exploitability. 

\item[Packages:] 
Successful exploitation depends on the existing packages installed in the target environment.
Removing these packages can prevent the attacker from exploiting a vulnerability. 
For example, in a typical remote exploitation scenario, the attacker initiates a remote shell connection to the target environment, which listens for such connections.
The $netcat$ package is the most common method used by attackers to open a listening socket in the target environment.
Therefore, removing the $netcat$ package can prevent such exploitations.
\end{description}
Considering all of the environmental conditions is impractical, so identifying the set of environmental conditions necessary to exploit a system is a crucial task.

\section{\label{sec:prop_method}The Proposed Framework}
\begin{figure}
    \centering
    \includegraphics[width=0.45\textwidth]{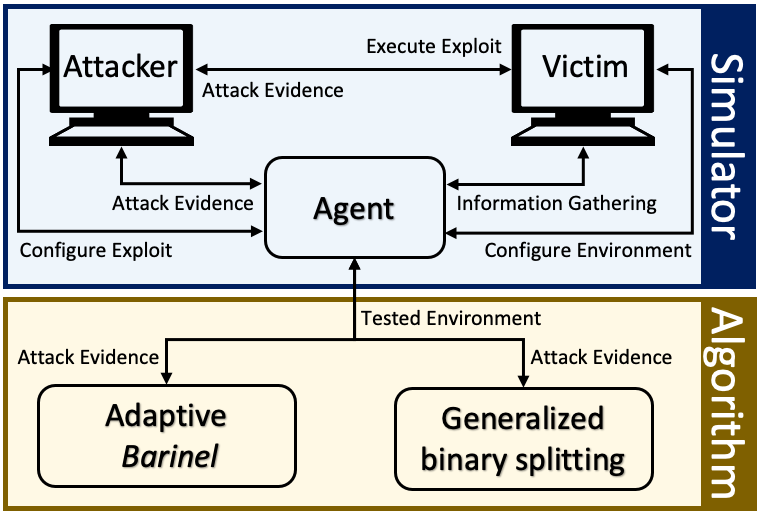} 
    \caption{The architecture of \methodname consists of two main components: a \textit{simulator} and an underlying \textit{algorithm}. The simulator automates the process of changing the system configuration, exploiting the system, and collecting evidence regarding the success of the exploitation. The algorithm is responsible for efficiently operating the simulator. Specifically, the algorithm determines the configuration of the environment to be tested by the simulator.
    \label{fig:framework}}
\end{figure}

In this section, we present our framework -- \methodname -- designed to solve the problem described in Section~\ref{sec:problem}.
The rational behind \methodname's design is that in order to identify the environmental conditions that affect the exploitability of a given security vulnerability, one must actually test the exploit on different environment configurations.
In addition, since modern environments consist of multiple configurable parameters that may influence the ability to exploit a vulnerability, manual testing cannot be considered.
Thus, a sophisticated testing strategy should be considered in order to minimize the number of tests required to solve the problem. 
To address these challenges, \methodname consists of two main components: a \textit{simulator} and an underlying \textit{algorithm} (see Figure~\ref{fig:framework}).

\subsection{\label{sec:arch}The Simulator}
The simulator we developed is a virtual environment implemented using Docker containers \cite{10.5555/2600239.2600241}.
The simulator consists of three main modules (see Figure~\ref{fig:framework}):

\begin{description}[style=unboxed,leftmargin=0cm]

\item[Victim (target environment).]
A Docker container installed with a vulnerable application that is remotely exploitable (i.e., remote attack vector).
The Victim is accessible through the network to both the Attacker and the Agent.
Our implementation of the simulator supports Linux-based victims.
However, the simulator can be easily extended to other platforms (e.g., Windows OS).

\smallskip

\item[Attacker.]
The attacking environment is responsible for attacking the Victim and collecting evidence with respect to the success of the exploitation. 
In order to automate the process of exploiting the target environment, the Attacker is implemented via a Metasploit-framework Docker container.
The Metasploit-framework~\cite{kennedy2011metasploit} is an open-source penetration testing tool developed by Rapid7.
The main reason for selecting the Metasploit-framework as an attacking environment is its support in the remote procedure call (RPC), which enables pen testers to automate the exploitation process. 
Exploiting a system using the Metasploit-framework is relatively simple and includes four main steps:
\begin{enumerate}[leftmargin=1.5\parindent]
        \item \textbf{Choosing and configuring an exploit:}
        In this step, the pen tester can select an exploit from the exploit database, which includes about 2,000 exploits.
        Each exploit includes several configurable parameters, such as the IP and port of the target machine, username, and password (if the exploitation requires authentication), etc.

        \item \textbf{Choosing and configuring a payload:} In this step, the pen tester should select a payload that is supported by the selected exploit. Metasploit supports various types of payloads (e.g., remote TCP shell, reverse TCP shell), most of which are used to open a remote terminal with the target environment. 
        Each payload includes several configurable parameters, such as the Attacker IP and port (for reverse shell payloads).

        \item \textbf{Executing the exploit:} In this step, the pen tester executes the exploit.
        
        \item \textbf{Running post-exploitation tools:} In this step, the pen tester can execute post-exploitation scripts on the target machine.
        We used post-exploitation tools to validate the success of the exploitation.
        In our implementation, we validate that a new session was created between the Attacker environment and the target machine. 
\end{enumerate}

\smallskip

\item[Agent.] The Agent is responsible for controlling and orchestrating the main process of the simulator.
This process is presented in Figure~\ref{fig:flow_chart}:
    \begin{enumerate}[leftmargin=1.5\parindent]
        \item \textbf{Attacker Configuration:}
        In this step, the Agent chooses the relevant exploit and payload to exploit the Victim.
        The set of parameters used to configure the exploit and payloads is determined based on the IP of the Victim, the port of the vulnerable application, and the IP of the Attacker; all other parameters are set to their default values (as determined by Metasploit).
        
        \item \textbf{Information Gathering:}
        The main objective of this step is to determine the initial set of environmental conditions that are relevant for the specific Victim.
        In the current implementation of the simulator, the set of environmental conditions is determined based on the services running, open ports, packages installed, and file permissions in the $\textbackslash bin$ directory.
        
        \item \textbf{Test Selection:}
        In this step, the Agent sends the previous test result to the Algorithm component, which responds with a new set of environmental conditions that should be tested by the simulator.
        
        \item \textbf{Victim Configuration:}
        In this step, the Agent modifies the environment configuration of the Victim.   
        The modifications to the Victim machine are determined by the Algorithm.

        \item \textbf{Execute Exploit:}
        In this step, the Agent instructs the Attacker to (a) exploit the Victim, and (b) collect evidence with respect to the success of the exploitation.
    \end{enumerate} 

\tikzstyle{decision} = [diamond, draw, fill=white!20,text width=1cm, text height=0.3cm, text centered]

\tikzstyle{block} = [rectangle, draw, fill=white!20,text width=2cm,text height=0.3cm, text centered,minimum width=1.2cm, minimum height=1cm]

\tikzstyle{line} = [draw, -latex']

\tikzstyle{start} = [draw, circle,fill=green!20, node distance=5cm, text height=0.3cm, text width=1cm,text centered]

\tikzstyle{end} = [draw, circle,fill=red!20, node distance=3cm,text height=0.3cm, text width=1cm,text centered]

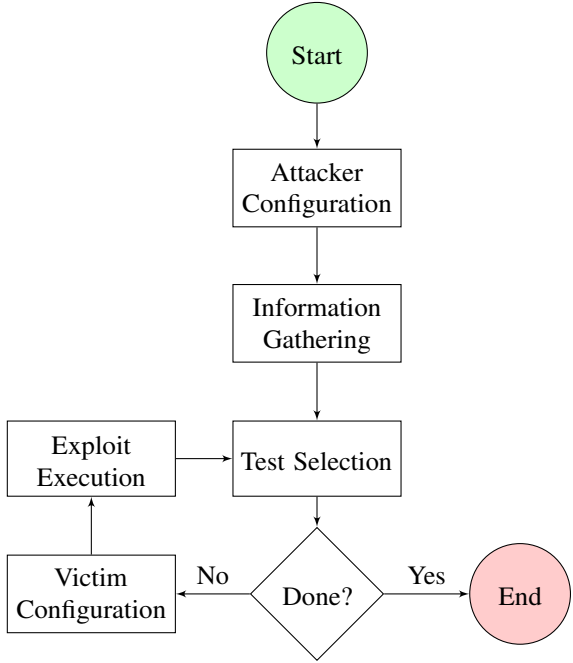
\begin{figure}
\centering   
\begin{tikzpicture}[node distance = 1.8cm, auto]
    '

    \node [start] (start) {Start};
    \node [block, below of=start] (init) {Attacker Configuration};
    \node [block, below of=init] (identify) {Information Gathering};
    \node [block, below of=identify] (evaluate) {Test Selection};
    \node [decision, below of=evaluate] (decide) {Done?};
    \node [block, left of=decide, node distance=3cm] (update) {Victim Configuration};
    \node [block, left of=evaluate, node distance=3cm] (execute) {Exploit Execution};
    \node [end, right of=decide, node distance=2.7cm] (end) {End};

    \path [line] (start) -- (init);
    \path [line] (init) -- (identify);
    \path [line] (identify) -- (evaluate);
    \path [line] (evaluate) -- (decide);
    \path [line] (decide) -- node[above] {No} (update);
    \path [line] (update) -- (execute);
    \path [line] (execute) -- (evaluate);
    \path [line] (decide) -- node[above] {Yes}(end);
\end{tikzpicture}
\caption{A flowchart that outlines the main process.}
\label{fig:flow_chart}
\end{figure}
\end{description}

\subsection{Algorithm}
The algorithm is responsible for efficiently operating the simulator.
Specifically, given a set of previously executed tests and their results, the Algorithm determines the specification of the next test to be executed by the simulator (i.e., the configuration of the Victim machine for the next test).
Note that in our implemented framework each iteration with the simulator (i.e., test) takes approximately 45 seconds; this demonstrates the importance of operating the simulator efficiently for a practical implementation.

As mentioned in definition~\ref{def:test}, we distinguish between testing in noiseless and noisy environments.

\subsection{\label{sec:gt}Testing in noiseless environments}
When testing in noiseless environments, the result of a test is deterministic.
That is, the result of a test is positive (equal to $1$) if $\exists j: r_j=e_j=1$ and negative (equal to $0$) otherwise. 
In order to solve the problem, we utilize the generalized binary splitting algorithm, a combinatorial, adaptive, and noiseless group testing algorithm.

\begin{description}[style=unboxed,leftmargin=0cm]
\item[Generalized binary splitting.] The algorithm receives two inputs (see  Algorithm~\ref{alg:generalized-binary-splitting}): a set of $n$ \textit{items} (denoted by $S$) and an upper bound for the number of \textit{defective items} (denoted by $\hat{d}$). 
The output of the algorithm is a set of defective items (denoted by $D$).
The algorithm starts by checking whether the number of remaining items is relatively close to the number of remaining defective items (i.e., $n \leq 2d-2$).
If yes, the algorithm tests each item individually and returns the current set of defective items (lines 4-7).
Otherwise, the algorithm tests a group of size $2^\alpha$ (where $\alpha = \log{\frac{n-\hat{d}+1}{\hat{d}}}$);
if the outcome is positive, it applies binary splitting to identify one defective item and $i$ non-defective items (lines 8-24).  
This procedure repeats until either $n$ or $\hat{d}$ are equal to zero.

\begin{algorithm}[t]
\SetAlgoLined

\SetKwInOut{Input}{Input}
\SetKwInOut{Output}{Output}

\Input{$S$ set of items}
\Input{$\hat{d}$ upper bound for the number of defective items}
\Output{$D$ set of defective items}

$D\leftarrow \emptyset$

$n\leftarrow |S|$

\While{$\hat{d} >$ 0 and $n > 0$}
{
\If{$n \leq 2\hat{d}-2$}
    {
        $D\leftarrow D \cup IndividualTesting(S)$
        
        break
    }
\Else
    {
        $l \leftarrow n-\hat{d}+1$
        
        $\alpha \leftarrow \lfloor \log_2(l/\hat{d}) \rfloor$.
        
        $subgroup = GetSubgroup(S,2^\alpha)$
        
        $test$\_$result \leftarrow Test(subgroup)$
        
        \If{($test$\_$result = 0$)} 
            {
                \tcp{test is negative}
                
                $n \leftarrow n-2^\alpha$
                
                $S \leftarrow S \setminus subgroup$
            }
        \Else
            {
                \tcp{test is positive}
                
                $item,idx \leftarrow BinarySplitting(subgroup)$
                
                $n \leftarrow n-1-idx$
                
                $\hat{d} \leftarrow \hat{d}-1$
                
                $D\leftarrow D \cup \{item\}$
                
                $S \leftarrow S \setminus GetSubgroup(S,1+idx)$

            }
    }
}

$return(D)$
 \caption{\label{alg:generalized-binary-splitting} Generalized binary splitting}
\end{algorithm}

\item[Adapting our problem to group testing.] In our modeling, \textit{environmental conditions} (denoted by $C=\{c_1,...,c_n\}$) are equivalent to \textit{items}, while the \textit{defective} items are equivalent to the environmental conditions that are \textit{necessary} for the exploitation (denoted by $R=\{r_1,...,r_n\}$).
Accordingly, $\hat{d}$ is an upper bound on the number of environmental conditions that are \textit{necessary} for the exploitation.

Figure~\ref{fig:gbs} illustrates the application of the generalized binary splitting algorithm for finding the environmental conditions that are necessary for the exploitation.
As can be observed, the algorithm starts by testing the first $2^{\alpha_{0}}$ environmental conditions ((1)).
Since not all of them are necessary for the exploitation, the test yields a negative result.
Then, the algorithm tests the next $2^{\alpha_{1}}$ environmental conditions ((2)).
Since a subset of them are required for the exploitation, the test yields a positive result.
Next, the algorithm applies binary splitting ((3)-(5)) to identify a specific environment condition that is necessary for the exploitation.
It should be noted that during binary splitting not all of the subgroups are tested.
Therefore, some environmental conditions may be classified as \textit{may be necessary} for the exploitation. 
These types of environmental conditions will be tested again as the search continues.

\begin{figure}[t]
    \centering
    \includegraphics[width=0.45\textwidth]{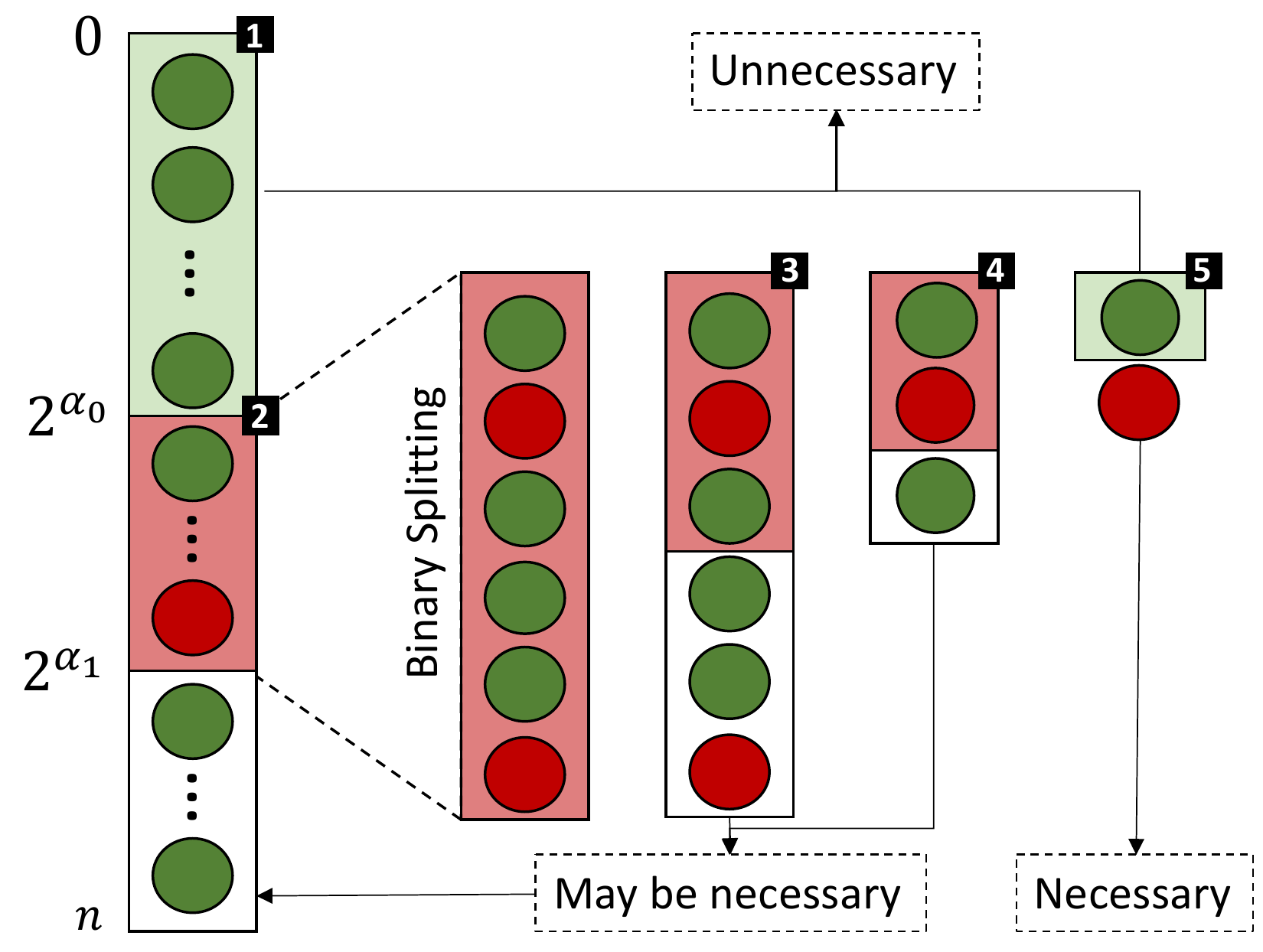}
    \caption{The application of generalized binary splitting for finding the environmental conditions that are necessary for the exploitation.
    Red/green circles represent environmental conditions that are necessary/unnecessary for the exploitation respectively. 
    Red/green rectangles represent positive/negative tests results.}
    \label{fig:gbs}
\end{figure}

\item[Time complexity.]
Given that the general number of environmental conditions (denoted by $n$) is sufficiently larger than the number of environmental conditions that are necessary for the exploitation, generalized binary splitting can find all of the defective items with $\hat{d}\cdot log(n/\hat{d})$ tests.
This dramatically improves the naive use of binary splitting to find all defective items (i.e., applying binary splitting $d$ times), which requires $O(d\cdot \log{n})$ tests.

\item[Practical considerations.]
As mentioned above, generalized binary splitting receives two inputs: a set of $n$ \textit{items} (denoted by $S$) and an upper bound for the number of \textit{defective items} (denoted by $\hat{d}$).
However, knowing (in advance) the exact number of environmental conditions that are necessary for exploiting the system is not trivial.
Therefore, in practice, $d$ is often estimated or an upper bound.
When $\hat{d}$ is overestimated, generalized binary splitting will identify all environmental conditions that are necessary for the exploitation however not as efficiently as when the correct value of $d$ is assigned.  
On the other hand, when $\hat{d}$ is underestimated, generalized binary splitting will identify up to $\hat{d}$ environmental conditions that are necessary for the exploitation, which may result in false negatives.
The common approach for identifying that $\hat{d}$ has been underestimated is to apply a test on all of the remaining environmental conditions. 
If the result is positive, then $\hat{d}$ was underestimated.
In that case, one would apply the generalized binary splitting algorithm on the remaining environmental conditions, while assigning a value of $d'$ to represent the number of environmental conditions that are necessary for the exploitation (among the remaining environmental conditions).
\end{description}

\subsection{\label{sec:br}Testing in noisy environments}
When testing in noisy environments, the result of a test is not deterministic. 
That is, the result of a test may be negative (equal to $0$) with some probability $\epsilon$ even when $\exists j: r_j=e_j=1$.
The reason for this non-deterministic behavior is because a successful change of an environmental condition often depends on additional parameters which are not considered during the test.
For instance, the deletion of a file may fail, if the file is currently under use by some process;
similarly, the deletion of a user may fail, if the user is running some process.
In order to address the non-deterministic (noisy) scenario, multiple noise models have been presented in the group testing literature \cite{aldridge2019group}.
In this work, we consider a specific variant of the dilution noise model, which addresses the type of non-determinism mentioned above.

The dilution noise model for a test containing $m$ items, $l$ of which are defectives, is provided by the following probability transition function:
\begin{equation}
\begin{split}
    &P(test=positive|m,l)  = 1 - \epsilon^l \\
    &P(test=negative|m,l)  = \epsilon^l
\end{split}
\label{eq:dilution_noise}
\end{equation}
Where $\epsilon$ is a noise parameter that estimates the probability for a defective item to produce a correct result.
The main drawback of this noise model, is that it cannot model environments such that each environmental condition that is necessary for the exploitation has different probability to produce negative result.
Therefore, in our application we consider the following more realistic noise model:  

\begin{equation}
\begin{split}
    &P(test=positive|m,l)  = 1 - \prod_{{c_i} \in l}^{} \epsilon_{c_i} \\
    &P(test=negative|m,l)  = \prod_{{c_i} \in l}^{} \epsilon_{c_i}
\end{split}
\label{eq:dilution_noise}
\end{equation}

Where $\epsilon_{c_i}$ is a noise parameter that estimates the probability for an environmental condition $c_i$ to produce a negative result.
We are not familiar with any literature of noisy adaptive group testing that can handle such noise model.
In order to solve this problem, we developed a variant of the Barinel~\cite{abreu2009spectrum} algorithm (coined Adaptive Barinel) that can be used to address this specific noisy adaptive group testing setup.

\begin{figure}[t]
\begin{subfigure}{.45\textwidth}
  \centering
  \includegraphics[width=0.95\textwidth]{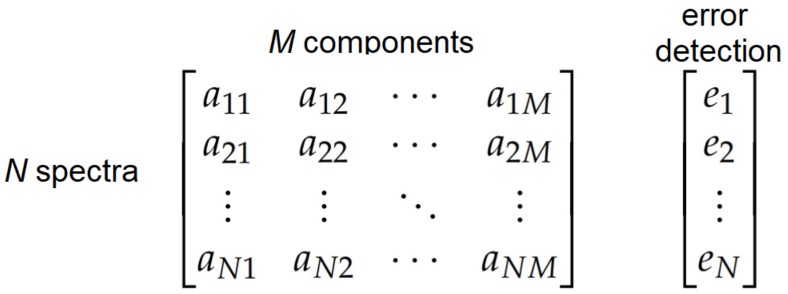}
  \caption{Original input as presented in\cite{abreu2009spectrum}}
  \label{fig:regular_barinel}
\end{subfigure}
\begin{subfigure}{.45\textwidth}
  \centering
  \includegraphics[width=1\textwidth]{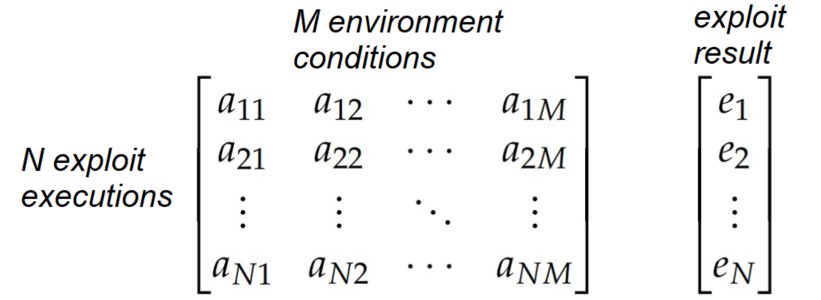}
    \caption{The inputs given our modeling}
    \label{fig:ours_barinel}
\end{subfigure}
\caption{Inputs to the Barinel algorithm}
\label{fig:fig}
\end{figure}

\begin{description}[style=unboxed,leftmargin=0cm]
\item[Barinel.] Barinel is an automated fault localization technique specifically designed for software diagnosis.
The key insight used by Barinel (presented in Algorithm~\ref{alg:barinel}) is that if a test fails, we can infer that at least one of the components (e.g. functions) in its trace is faulty. 
Thus, the trace of a failed test is a conflict \cite{stern2012exploring}, which includes the components  that participated the trace. 
Obviously, since the test failed, at least one of the  components in the conflict is faulty. 
A hitting set (i.e., minimal set cover \cite{cormen2009introduction}) of these failed tests is a diagnosis \cite{de1987diagnosing}, where a diagnosis is a set of faulty components that caused the tests to fail.  
Barinel adopts Staccato~\cite{abreu2009low} to generate the hitting sets.

Barinel receives two inputs (see Figure~\ref{fig:regular_barinel}): An activity matrix (denoted by $A$) of a size $N \times M$, which describes the presence or absence of each of the $M$ components in the traces of each of the $N$ tests (having $A_{i,j}=1$ indicates that component $j$ is in the trace of test $i$); and an error vector (denoted by $e$) of a size $N$, which represents the corresponding test results (success/fail).
These two inputs together are referred to as observations (denoted by $OBS$).
The Barinel's output  is a diagnostic report, which includes a list of diagnoses and the likelihood for those diagnoses to be correct.
In order to produce the diagnostic report (line 11), Barinel follows a Bayesian approach:

\begin{equation}
    P(d|OBS) = \frac{P(OBS|d) \cdot P(d)}{P(OBS)}
    \label{eq:ranking}
\end{equation}

\noindent That is, the likelihood for a diagnosis $d$ to be correct is an estimate of the posterior probability that $d$ is the correct diagnosis given the observations $OBS$, where $P(d)$ is the prior probability of $d$ being correct (i.e., representing the faulty components) and $P(OBS)$ is a normalization factor. 

To compute $P(OBS|d)$ (line 3), Barinel assumes that the tests are independent and therefore can be computed as follows:

\begin{equation}
  P(OBS|d) = \displaystyle\prod_{obs_i \in OBS}^{} P(obs_i|d)  
\end{equation}

\noindent where $obs_i$ is a single observation, and the probability $P(obs_i|d)$ is calculated as follows:

\begin{equation}
P(obs_i|d) \equiv
\begin{dcases*}
\displaystyle\prod_{j\in d \wedge obs_i}^{} g_j
   & if $obs_i$ passed\ \\[1ex]
1-\displaystyle\prod_{j\in d \wedge obs_i}^{} g_j
   & if $obs_i$ failed\
\end{dcases*}
\label{eq:old-goodness}
\end{equation}

\noindent where $g_j$ is a goodness factor that describes the probability that a faulty component $j$ produces correct output. \\
Barinel estimates these goodness factors by solving an optimization problem in which the goodness factors are the variables that are set so as to maximize $P(OBS|d)$, using a gradient ascent technique (line 9).

\item[Adaptive version of Barinel.]
The textbook implementation of Barinel cannot be used for our purposes for the following reasons.
First, Barinel receives all of the observations in advance, in contrast to our adaptive framework which uses the results of previous tests to determine the next test to be executed. 
Second, Barinel's objective is to generate a diagnostic report that explains the observations.
In contrast, our objective is to intelligently select the next tests to be executed, while minimizing the total number of tests overall.
In order to address these differences, we developed an adaptive version of Barinel (coined Adaptive Barinel).

One of the first challenges we faced when developing the adaptive version of Barinel is very similar to the exploration-exploitation trade-off, which is a well-known dilemma in the fields of decision-making, recommendation systems, and reinforcement learning \cite{sutton2018reinforcement}.
Traditionally, the exploration-exploitation trade-off confronts two contradicting approaches in order to maximize the total gain (also known as the expected sum of rewards):
Exploitation \textemdash~repeat decisions/actions that have already operated properly in order to gain an expected reward; and Exploration \textemdash~ explore novel decisions/actions, while expecting to gain a greater reward.
We faced a similar dilemma: Selecting a test that analyzes the components that were already suspected as faulty, in order to further confirm this suspicion, i.e., reduce false positives (which we refer to as \textit{exploitation});
or, selecting a test that explores new components, in order to identify more components that may be faulty, i.e., increase the number of true positives (which we refer to as \textit{exploration}).
To overcome this challenge, we were inspired by the $\epsilon-greedy$ strategy~\cite{watkins1989learning}, which represents a balance between exploration and exploitation.

Adaptive Barinel is presented in Algorithm~\ref{alg:adaptive_barinel}.
The algorithm receives the following inputs: the set of components to be considered during the test (denoted by $M$), an estimate of the number of faulty components (denoted by $\hat{d}$), the exploration ratio (denoted by $\epsilon$), the decay factor (denoted by $decay$), the minimal exploration ratio (denoted by $\epsilon_{min}$), the bootstrapping length (denoted by $b$), and the decoding frequency (denoted by $f$).
The output of the algorithm is a set of components and the likelihood of them to be faulty (denoted by $P$).
The algorithm starts by calculating $2^\alpha$, which determines the number of components that will be considered in each test (line 1); this is done similar to the generalized binary splitting algorithm.
Then, the algorithm randomly samples $b$ tests, where each test considers $2^\alpha$ components (that is, in each row in $A$ only $2^\alpha$ indexes are set to $1$); and executes those tests to receive the corresponding error vector (lines 3-4).
We refer to this phase as \textit{bootstrapping}, since the algorithm only performs \textit{exploration}. 
Next, the algorithm applies Barinel to receive a diagnostic report.
Since the diagnostic report will probably not change dramatically in each iteration and the application of Barinel is time-consuming, we apply Barinel every $f$ iterations.
Given the diagnostic report, we calculate the likelihood of each component to be faulty (line 8).
This is done according to Equation~\ref{eq:sum}.
\begin{equation}
    P(c) = \sum_{d\in D}^{}
    \begin{cases}
      P(d), & c\in d \\
      0, & c\not\in d
    \end{cases}
\label{eq:sum}
\end{equation}
Based on the diagnostic report, we select $\hat{d}$ components that have highest probability (line 9).
These components will be used by the $\epsilon-greedy$ strategy.
Specifically, in the \textit{exploration} phase, we randomly sample a single test by considering $2^\alpha$ components from $M-T$ (line 12).
In the \textit{exploitation} phase, we sample a single test by considering a single component from $T$ (line 15).
The rational behind this sampling technique is to use group testing to accelerate exploration and individual testing to reduce the number of false positives.

\item[Adapting our problem to software fault diagnosis.] In our modeling (see Figure \ref{fig:ours_barinel}), \textit{environmental conditions} (denoted by $C=\{c_1,...,c_n\}$) are equivalent to the tested components, while a \textit{faulty component} (denoted by $\neg h$) is equivalent to an environment condition that is \textit{necessary} for the exploitation (denoted by $R=\{r_1,...,r_n\}$). 
Similarly, a \textit{tested environment} (denoted by $E=\{e_1,...,e_n\}$) is equivalent to a trace.\\
Specifically, a tested component $A_{i,j}=1$, if during the $i^{th}$ test, the \textit{environment condition} $c_j$ was enabled ($c_j=1$), and vice-versa.
Correspondingly, within the error vector $e_i=1$ if the tested environment $E$ is not exploitable, and vice versa.

\item[Time complexity.]
A theoretical lower bound is not yet to be presented for the dilution noise model.
Furthermore, when the algorithm is based on sampling, theoretical analysis is very challenging.
Therefore, in practice, such algorithms are mainly evaluated empirically through simulations.
An empirical evaluation of the Adaptive Barinel algorithm is presented in Section \ref{sec:eval_noisy}.

In addition to sampling, our algorithm uses Barinel to generate diagnostic.
Here we evaluate the time complexity for this operation.
Barinel includes three main (computationally heavy) procedures:
\begin{enumerate}
    \item \textbf{Generating the set of diagnoses:} The set of diagnoses, which is defined as the hitting sets over the traces of failed tests, is calculated using the \textit{Stoccato} algorithm.
    The time complexity of \textit{Stoccato} is estimated to be $O(N \cdot M)$.
    \item \textbf{Health probability estimation:} The health probability, which determines the likelihood that a faulty component generates a correct result, is calculated using maximum likelihood estimation (MLE) via gradient ascent.
    The time complexity of the MLE is independent of the size of $N$ and $M$, yielding a constant time complexity (denoted by $C$).
    \item \textbf{Sorting the diagnostics in the report:} The report is ordered in descending order.
    It can be assumed that for large systems $|D|$ would scale with $M$. Therefore, the time complexity for sorting the report is  $M \cdot log(M)$.
\end{enumerate}

Thus, given an activity matrix of size  $N \times M$, Barinel's complexity is $O(M \cdot N + M \cdot log(M))$.
It should be mentioned that it is reasonable to assume that much less effort is required to run Barinel than to perform a test.
\item[Practical considerations.]
As mentioned above, Adaptive Barinel receives the following seven inputs: $M$, $\hat{d}$, $\epsilon$, $decay$, $\epsilon_{min}$, $b$, and $f$.
In this section, we discuss the considerations for determining these parameters.
Based on our experiments, the balance between exploration and exploitation is crucial for an efficient search.
Specifically, on the one hand, insufficient exploration will result in low recall.
However, on the other hand, insufficient exploitation will result in an inefficient search.
The parameters that are responsible for balancing exploration and exploitation are  $\epsilon$, $decay$, $\epsilon_{min}$, and $b$.

Another important parameter is the decoding frequency ($f$). 
Naturally, a high decoding frequency will result in a more accurate assessment.
However, it can be very time-consuming and unnecessary, since the diagnostic report probably won't dramatically change in each iteration.

Last but not least is the $\hat{d}$  parameter, which is an estimator for the number of faulty components. 
Based on this parameter, the algorithm determines the number of components included in each test.
Concretely, on the one hand, selecting $\hat{d}$ such that $\frac{n}{\hat{d}}$ is relatively small will result in tests with a large number of components.
On the other hand, selecting $\hat{d}$ such that $\frac{n}{\hat{d}}$ is relatively large will result in tests with a large number of components.The intuition behind this parameter is that if $\frac{n}{\hat{d}}$ is relatively small, then the probability that a faulty component will be included in a test is also small.
Therefore, we can accelerate the search by conducting tests with a large number of components. 
On the other hand, if $\frac{n}{\hat{d}}$ is relatively large, then the probability that a faulty component will be included in a test is also large.
Therefore, tests with a small number of components should be considered. 
\end{description}

\begin{algorithm}[h]
\SetAlgoLined
\SetKwInOut{Input}{Input}
\SetKwInOut{Output}{Output}

\Input{$M$ set of components}
\Input{$\hat{d}$ an estimate for the number of faulty components}
\Input{$\epsilon$ exploration ratio}
\Input{$decay$ decay factor}
\Input{$\epsilon_{min}$ minimal exploration ratio}
\Input{$b$ bootstrapping length}
\Input{$f$ decoding frequency}
\Output{$P$ set of components and the likelihood of them to be faulty}

$\alpha \leftarrow \lfloor \log_2(\frac{n-\hat{d}+1}{\hat{d}}) \rfloor$.

$i \leftarrow 0$

\tcp{bootstrapping}

$A \leftarrow$  RandomSample($M$,$2^\alpha$,[|M|,b])

$e \leftarrow Test(A)$

\Repeat{Convergence criteria is met}{

    \If{$i \% f == 0$}
        {
                    D $\leftarrow$ Barinel(A,e)
                    
                    P $\leftarrow$ ProbabilitySummation(D)
                    
                    T $\leftarrow$ top$(\hat{d},P)$
        }
    
    \If{\upshape{RandomSample}$([0, 1]) <  \epsilon$}
        {
            \tcp{exploration}
            $test \leftarrow $ RandomSample ($M-T$,$2^\alpha$,[|M|,1])
        }
    \Else
       {
       
       \tcp{exploitation}
       
       $test \leftarrow $ RandomSample ($T$,$1$,[|M|,1])
       }
       
        e $\leftarrow$ e $\cup ~$ Test~($test$)
        
        A $\leftarrow$ A $\cup ~ test$
        
        $i \leftarrow i+1$

        $\epsilon \leftarrow max(\epsilon_{min},\epsilon * decay)$
        
}

return($P$)

 \caption{\label{alg:adaptive_barinel} Adaptive Barinel}
\end{algorithm}

\begin{algorithm}[h]
\SetAlgoLined

\SetKwInOut{Input}{Input}
\SetKwInOut{Output}{Output}

\Input{$A$ Activity Matrix}
\Input{$e$ Error Vector}
\Output{$D$ Diagnostic report}

$D$  $\leftarrow$ Staccato((A, e))

\For{$d$ $\in~D$}{
$prob$ $\leftarrow$ GeneratePr((A, e), dk)

$idx$ $\leftarrow$ 0

$Pr[d][idx]$ $\leftarrow$ 0

\Repeat{$|Pr[d][idx]-Pr[d][idx-1]| \leq \epsilon$}{
        
        $idx$ $\leftarrow$ $idx+1$
        
        \For{$j$ $\in~d$}{
            $g_j\leftarrow g_j + \alpha \cdot \nabla prob[g_j]$
            
        }
        $Pr[d][idx] \leftarrow$ Evaluate$(prob,\forall_{j\in d_k} g_j)$
        
    }
}

return sort($D$, Pr)
 \caption{\label{alg:barinel} Barinel}

\end{algorithm}

\section{\label{sec:evaluation}Evaluation}

\subsection{\label{subsec:vul} Vulnerabilities and exploits tested}
We evaluated \methodname using the following vulnerabilities and exploits:
\begin{description}[style=unboxed,leftmargin=0cm]

\item[\textbf{CVE-2007-2447}:]
Improper input validation in the MS-RPC functionality in \textit{Samba server} versions 3.0.0 - 3.0.25rc3 allows remote attackers to execute arbitrary code via remote injection of shell commands. 

CVSS classifies this vulnerability as medium severity, with a base score of 6.0, impact score of 6.4, and exploitability score of 6.8.
The reason for the medium exploitability score is because the exploitation can be performed with medium complexity (i.e., the vulnerability can only be exploited in specific configurations, which are not the default) over the network and requires a single authentication (i.e., exploiting the vulnerability requires one instance of authentication).

This vulnerability can be exploited using the \textit{exploit/multi/samba/usermap\_script} Metasploit module.
This module exploits the vulnerability by initiating an anonymous call with Linux shell commands to the \textit{SamrChangePassword()} MS-RPC function (in combination with the "username map script" option) 
The unescaped shell commands are passed as arguments to \textit{/bin/sh} allowing for remote command execution.

\item[\textbf{CVE-2010-2075}:]
Backdoor in \textit{UnrealIRCd} version 3.2.8.1 allows remote attackers to execute arbitrary code.

CVSS classifies this vulnerability as high severity, with a base score of 7.5, impact score of 6.4, and exploitability score of 10.
The reason for the high exploitability score is because the exploitation can be performed with low complexity, over the network, without any authentication.

This vulnerability can be exploited using the \textit{exploit/unix/irc/unreal\_ircd\_3281\_backdoor} Metasploit module.
This module exploits the vulnerability by opening a socket to the Victim machine and sending the backdoor command \textemdash~ the string "AB" followed by Linux shell commands.
Those commands are handed off directly to \textit{system()}, a C library function that passes the command to the host environment to be executed by the command processor.

\item[\textbf{CVE-2011-2523}:] Backdoor in \textit{vsftpd} version 2.3.4 allows remote attackers to open shell on port 6200.

CVSS classifies this vulnerability as critical, with a base score of 10.0, impact score of 10.0, and exploitability score of 10.0.
The reason for the high exploitability score is because the exploitation can be performed with low complexity, over the network, without any authentication.

This vulnerability can be exploited using the \textit{exploit/unix/ftp/vsftpd\_234\_backdoor} Metasploit module which adds a ":)" smiley face into the username that triggers the machine to open a shell on port 6200 by executing \textit{/bin/sh}.

\item[\textbf{CVE-2011-3556}:] Misconfiguration in the Java Runtime Environment in \textit{Java SE JDK and JRE} version 7 allows remote attackers to execute arbitrary code. 

CVSS classifies this vulnerability as high severity, with a base score of 7.5, impact score of 6.4, and exploitability score of 10.0.
The reason for the high exploitability score is because the exploitation can be performed with low complexity, over the network, without any authentication.

This vulnerability can be exploited using the \textit{exploit/multi/misc/java\_rmi\_server} Metasploit module.
The exploit takes advantage of the RMI distributed garbage collector available from every RMI endpoint, which allows loading classes from any remote (HTTP) URL.
\end{description}

\subsection{\label{sec:ev_setup}Evaluation setup}
\begin{description}[style=unboxed,leftmargin=0cm]

\item[Victim machine:]
A docker container running Linux Ubuntu 8.0.4.
The container is based on Metasploitable 2~\cite{moore2012metasploitable}, an intentionally vulnerable virtual machine, which is exposed to the vulnerabilities described in Section~\ref{subsec:vul}.

\item[Agent implementation:]
In our implementation of the Agent (which is based on the DeepExploit~\cite{DeepExploit} tool) the following actions are supported: 
\begin{enumerate}
    \item \textbf{Services:} Stopping a service  using the \textit{etc/init.d stop service\_name} utility.
    \item \textbf{Packages:} Deleting a package using the \textit{dpkg -r package\_name} utility. 
    \item \textbf{Connectivity:} Changing firewall rules, i.e., blocking network ports using the \textit{iptables port INPUT/OUTPUT -p PROTOCOL --destination-port port -j DROP} utility.
    \item \textbf{Access control:} Changing file permissions using the \textit{chmod} utility, e.g., restricting the access to files in $\textbackslash bin$ directory.
\end{enumerate}

It should be mentioned that our Agent implementation can be extended to support additional actions.
Given those three actions, the number of environmental conditions that were considered in the evaluation is 675. 
Since some actions may affect the operation of the simulator, we created a \textit{Blacklist} that prevents the execution of some predefined actions. 
For instance, the simulator uses the \textit{dpkg} package to delete packages, thus removing the \textit{dpkg} package will affect the operation of the simulator.
For this reason, we put such actions (\textit{dpkg -r dpkg}) in the \textit{Blacklist}.
After removing blacklisted actions, the number of environmental conditions considered in the evaluation was decreased to 642.

\item[System setup:]
The evaluation was conducted on a 64-bit Windows server 2008 R2 Enterprise machine, with a  2.00 GHZ Intel Xeon CPU (version E5-2620, 24 logical cores) and 64 GB of RAM.
The entire framework is implemented in Python. 
Specifically, the simulator and generalized binary splitting algorithm were implemented in Python 3.6, and the Adaptive Barinel algorithm was implemented in Python 2.7.
The communication of the Agent and the Attacker docker container is performed via the RPC protocol.
The communication of the Agent and the Victim container is performed via the docker SDK for Python.
The communication of the Agent and the Algorithms is performed via TCP sockets.
\end{description}

\subsection{Testing in noiseless environments}
In this evaluation scenario, we assume that our environment is noiseless, i.e., the result of a test is positive (equal to $1$) if $\exists j: r_j=e_j=1$ and negative (equal to $0$) otherwise.
Given that, we selected the generalized binary splitting algorithm to operate the simulator.

The results of the assessment are presented below:
\begin{description}[style=unboxed,leftmargin=0cm]
\item[\textbf{CVE-2007-2447}:]
The proposed framework identified that a successfull exploitation of this vulnerability is rooted in the ability of the exploit to pass unescaped Linux shell commands as arguments to \textit{/bin/sh}.
In order to do so, the user running the vulnerable application (in this case the vulnerable samba daemon) must have privileges to run \textit{/bin/bash} (or \textit{/bin/sh}).
Thus, this vulnerability cannot be exploited in environments that restrict the \textit{execute} privilege of \textit{/bin/bash}, such that the user running the vulnerable application will not be able to open a shell.

In addition, the framework identified that the payloads available utilize two different approaches when compromising the system:
First, using third party libraries (\textit{telnet}, \textit{netcat}, \textit{openssl}, and \textit{socat}) to create an interactive shell from the Victim environment to the Attacker environment.
Second, using scripting languages, such as \textit{ruby} and \textit{perl}, to open a network socket between the Attacker and Victim machine and use it as a communication channel to a local terminal (i.e., by redirecting STDIN and STDOUT).
Thus, it is more difficult to exploit this vulnerability in environments that do not include netcat, OpenSSL, socat, Ruby, or Perl.

\item[\textbf{CVE-2010-2075}:]
The proposed framework identified that a successfull exploitation of this vulnerability is rooted in the ability of the exploit to pass Linux shell commands to the \textit{system()} C library function, which opens a shell and run the commands.
In order to do so, the user running the vulnerable application (in this case the vulnerable IRC daemon) must have privileges to run \textit{/bin/bash} (or \textit{/bin/sh}).
Thus, this vulnerability cannot be exploited in environments that restrict the \textit{execute} privilege of \textit{/bin/bash}, such that the user running the vulnerable application will not be able to open a shell.

In addition, the framework identified that the payloads available utilize two different approaches when compromising the system:
First, using \textit{telnet} to create an interactive shell from the Victim environment to the Attacker environment. 
Second using scripting languages, such as \textit{ruby} and \textit{perl}, to open a network socket between the Attacker and Victim machine and use it as a communication channel to a local terminal (i.e., by redirecting STDIN and STDOUT).
Thus, it is more difficult to exploit this vulnerability in environments that do not include \textit{telnet}, \textit{ruby}, or \textit{perl} packages.

\item[\textbf{CVE-2011-2523}:] 
The proposed framework identified that a successful exploitation of this vulnerability is rooted in the ability of the exploit to (1) open a listening socket on port 6200, and (2) use this socket as a communication channel between the Attacker machine to a local terminal (i.e., by redirecting STDIN and STDOUT) opened via a call to the \textit{execl()} C library function with the path \textit{/bin/sh} as a parameter.
Thus, this vulnerability cannot be exploited in environments that (1) restrict incoming connections on port 6200, or (2) that restrict the \textit{execute} privilege of \textit{/bin/bash}, such that the user running the vulnerable application (\textit{vsftpd} service in this case) will not be able to open a shell. 

In addition, the framework identified that by removing the \textit{libselinux1} package, the vulnerability cannot be exploited.
According to information on the Debian website,\footnote{\url{https://packages.debian.org/sid/libselinux1}} \textit{libselinux1} provides an API for SELinux applications for get and set process and file security contexts and to obtain security policy decisions.
Despite our in depth examination of this, we were unable to fully understand why removing that package prevents the exploitation.
The main insight from our analysis is that during the authentication process \textit{vsftpd} (indirectly) calls some SELinux API function, which does not exist when the \textit{libselinux1} package is removed.
Surprisingly, the \textit{vsftpd} application did not crash, but we assume that the source code of the backdoor is no longer reachable, thus preventing the exploitation.

\item[\textbf{CVE-2011-3556}:]
The proposed framework identified that a successful exploitation of this vulnerability is rooted in the ability of the exploit to access the Java RMI server (running on port 1099), which is enabled by default in Java SE JDK and JRE below version 7.
Thus, this vulnerability cannot be exploited in environments that restrict incoming connections on port 1099.
\end{description}

\begin{figure}[t]
    \centering
    \includegraphics[width=0.45\textwidth]{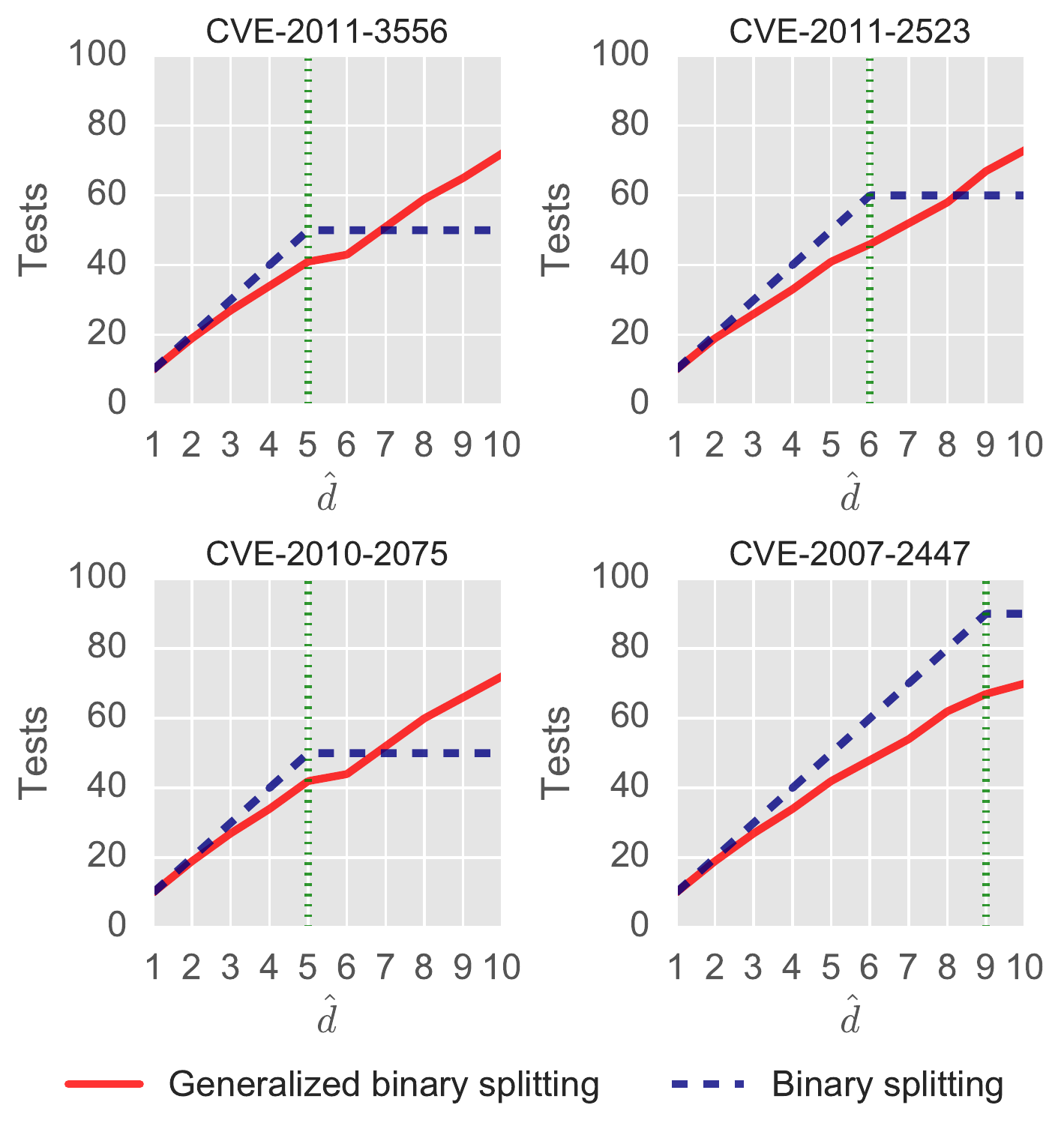} 
    \caption{The performance of the generalized binary splitting algorithm compared to the binary splitting algorithm at different ($\hat{d}$) upper bound values. The green curve represents the true number of environmental conditions that are necessary for exploitation of the vulnerability. }
    \label{fig:results_gbs}
\end{figure}

In Figure \ref{fig:results_gbs} we evaluate the performance of the generalized binary splitting algorithm, given different values of $\hat{d}$.
We compared the results with the alternative approach of applying the binary splitting algorithm $d$ times.
As can be seen, when $\hat{d}$ is relatively close to $d$, the generalized binary splitting algorithm produces better results.
However, when $\hat{d}$ largely overestimates $d$, the binary splitting algorithm produces better results.

\subsection{\label{sec:eval_noisy}Testing in noisy environments}
In this evaluation scenario, we assume that our environment is noisy (i.e., the result of a test may be negative (equal to $0$) with some probability $\epsilon$ even when $\exists j : r_j=e_j=1$).
In order to evaluate the performance of the Adaptive Barinel algorithm when dealing with noise, we conducted a controlled experiment in which we simulated noise at different volumes. 
For each environmental condition $c_i$,  we sampled the noise parameter $\epsilon_{c_i}$ from a Gaussian distribution, i.e., $\epsilon_{c_i} \sim |\mathcal{N}(\mu,\,\sigma^{2})|$, where $\mu$ and $\sigma$ are the parameters of a Gaussian distribution which determines the mean and standard deviation of the simulated noise volume.

In our experiments, we tested the following four noise distributions: $\mathcal{N}(\mu=0.05\ , \sigma =0.05),~ \mathcal{N}(\mu=0.1\ , \sigma =0.05),~\mathcal{N}(\mu=0.15\ , \sigma =0.05),~ \mathcal{N}(\mu=0.2\ , \sigma =0.05).$
For each noise distribution we tested the recall and precision of the algorithm for the task of identifying the environmental conditions that are necessary for the exploitation of the vulnerabilities described in Section~\ref{subsec:vul}.

\begin{figure}[h!]
    \centering
    \includegraphics[width=0.45\textwidth]{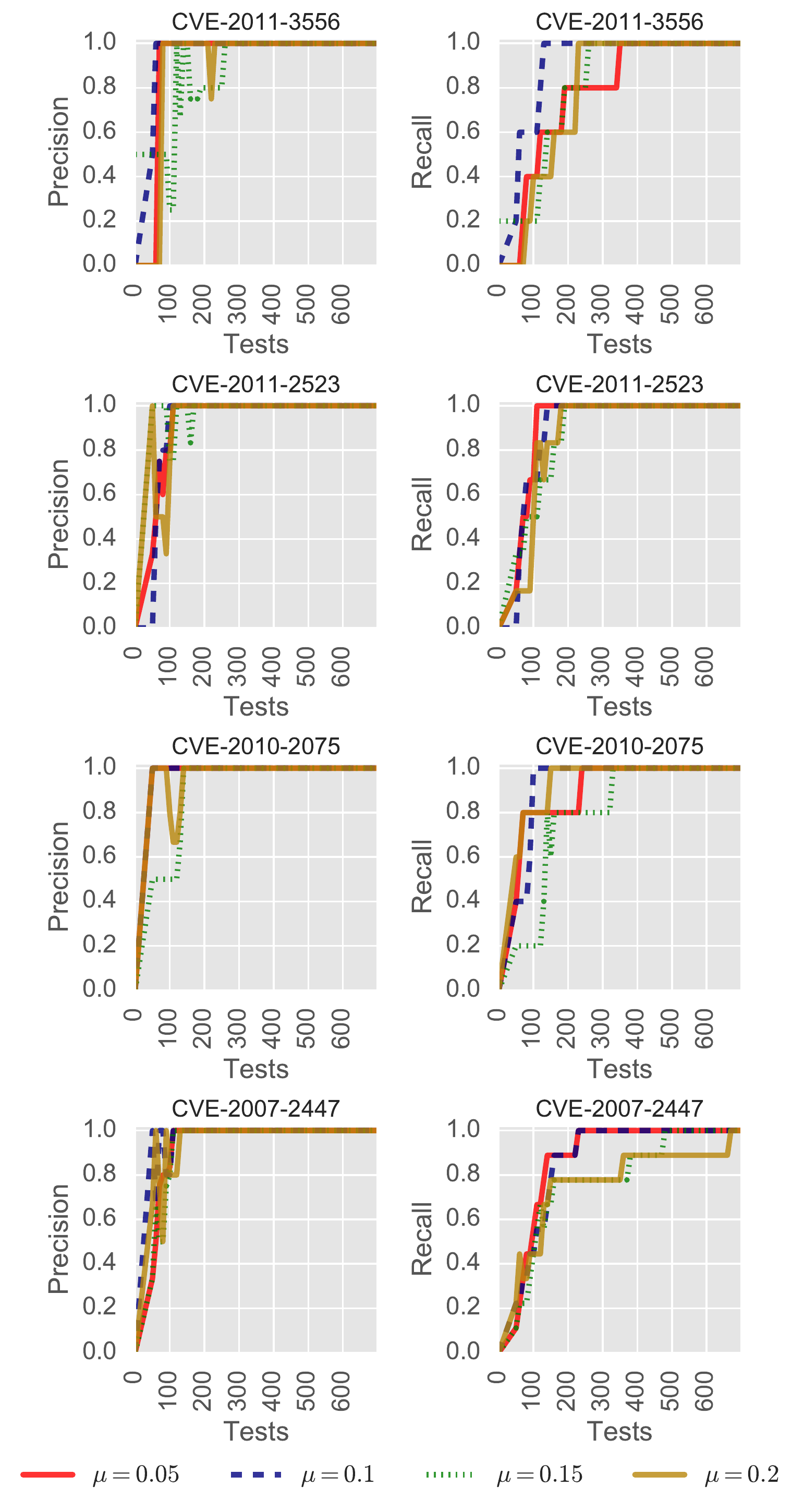} 
    \caption{The performance of the Adaptive Barinel algorithm in identifying the environmental conditions that are necessary for the exploitation of different vulnerabilities at different noise volumes (in terms of recall, precision, and number of tests). 
    }
    \label{fig:results_second}
\end{figure}

The results presented in Figure~\ref{fig:results_second} reveal very interesting insights.
First, the Adaptive Barinel algorithm is able to achieve perfect recall and precision at all noise distributions, however at high amount of noise, achieving such results takes more iterations.
For instance, when observing the performance of the algorithm on CVE-2007-2447, we can see that achieving a recall of one takes about 200 iterations when $\mu=0.05$ and about 600 iterations when $\mu=0.2$.  
Second, a very surprising insight is that the precision does not depend on the noise distribution. That is, even at high amount of noise, the algorithm does not produce false positives. 

\section{\label{sec:future_work}Conclusions and Future Work}
In this paper, we present \methodname, a novel framework and method for evaluating the exploitability of security vulnerabilities given a specific environment.
As part of this framework, we developed a simulator that automates the process of changing the system's configuration, exploiting the system, and collecting evidence with respect to the success of the exploitation.
In addition, we suggested two algorithms that can be used to efficiently operate the simulator in both noiseless and noisy environments.

Given a vulnerable environment and relevant exploit, \methodname can identify the specific environmental properties necessary for a successful exploitation.
With this knowledge, security analysts can evaluate the exploitability of their system by considering their environment's configuration, rather than by taking generic considerations (such as the exploitability metric presented in CVSS) into account, contributing to a more accurate risk management process.
Furthermore, these properties can indicate the specific hardening techniques required to reduce the attack surface. 

As future work we suggest the following research directions:

\begin{description}[style=unboxed,leftmargin=0cm]

\item[Extending the simulator:] The current implementation of the simulator supports four types of actions (see Section \ref{sec:ev_setup}).
Extending the simulator with additional actions, such as those presented in Appendix \ref{sec:appendix}, enables a more thorough analysis of the vulnerability and exploit.

\item[Utilizing prior knowledge:] The current implementation of the algorithms does not consider any prior knowledge on the vulnerability, exploit, or environmental conditions.
We believe that utilizing prior knowledge (e.g., using reinforcement learning algorithms) can accelerate the search.

\item[Additional environments:] The current implementation of the simulator supports Linux-based operating systems. We plan to extend the framework to support testing and evaluating the exploitability of Windows-based operating systems.
\end{description}

\bibliographystyle{plain}
\bibliography{bibliography/bibib.bib}

\begin{thebibliography}{10}

\bibitem{Nessuscite}
R. deraison, nessus, retrieved may 2003, from http://www.nessus.org, 1999.

\bibitem{abreu2009low}
Rui Abreu and Arjan~JC Van~Gemund.
\newblock A low-cost approximate minimal hitting set algorithm and its
  application to model-based diagnosis.
\newblock In {\em Eighth Symposium on Abstraction, Reformulation, and
  Approximation}, 2009.

\bibitem{abreu2009spectrum}
Rui Abreu, Peter Zoeteweij, and Arjan~JC Van~Gemund.
\newblock Spectrum-based multiple fault localization.
\newblock In {\em Proceedings of the 2009 IEEE/ACM International Conference on
  Automated Software Engineering}, pages 88--99. IEEE Computer Society, 2009.

\bibitem{Albanese2014ManipulatingTA}
Massimiliano Albanese, Ermanno Battista, Sushil Jajodia, and Valentina Casola.
\newblock Manipulating the attacker's view of a system's attack surface.
\newblock {\em 2014 IEEE Conference on Communications and Network Security},
  pages 472--480, 2014.

\bibitem{aldridge2019group}
Matthew Aldridge, Oliver Johnson, and Jonathan Scarlett.
\newblock Group testing: an information theory perspective.
\newblock {\em arXiv preprint arXiv:1902.06002}, 2019.

\bibitem{cormen2009introduction}
Thomas~H Cormen, Charles~E Leiserson, Ronald~L Rivest, and Clifford Stein.
\newblock {\em Introduction to algorithms}.
\newblock MIT press, 2009.

\bibitem{cowan1999protecting}
Crispin Cowan, Steve Beattie, Ryan~Finnin Day, Calton Pu, Perry Wagle, and Erik
  Walthinsen.
\newblock Protecting systems from stack smashing attacks with stackguard.
\newblock In {\em Linux Expo}, 1999.

\bibitem{de1987diagnosing}
Johan De~Kleer and Brian~C Williams.
\newblock Diagnosing multiple faults.
\newblock {\em Artificial intelligence}, 32(1):97--130, 1987.

\bibitem{denis2016penetration}
Matthew Denis, Carlos Zena, and Thaier Hayajneh.
\newblock Penetration testing: Concepts, attack methods, and defense
  strategies.
\newblock In {\em 2016 IEEE Long Island Systems, Applications and Technology
  Conference (LISAT)}, pages 1--6. IEEE, 2016.

\bibitem{developers2012open}
OpenVAS Developers.
\newblock The open vulnerability assessment system (openvas), 2012.

\bibitem{ding-zhudu1993}
Ding-Zhu Du.
\newblock {\em Combinatorial Group Testing and Applications (Applied
  Mathematics)}.
\newblock World Scientific Pub Co Inc, nov 1993.

\bibitem{hassell2006hardening}
Jonathan Hassell and Obis Orlando.
\newblock {\em Hardening Windows}.
\newblock Springer, 2006.

\bibitem{hwang1972method}
FK~Hwang.
\newblock A method for detecting all defective members in a population by group
  testing.
\newblock {\em Journal of the American Statistical Association},
  67(339):605--608, 1972.

\bibitem{kennedy2011metasploit}
David Kennedy, Jim O'gorman, Devon Kearns, and Mati Aharoni.
\newblock {\em Metasploit: the penetration tester's guide}.
\newblock No Starch Press, 2011.

\bibitem{landoll2005security}
Douglas~J Landoll and Douglas Landoll.
\newblock {\em The security risk assessment handbook: A complete guide for
  performing security risk assessments}.
\newblock CRC Press, 2005.

\bibitem{mell2006common}
Peter Mell, Karen Scarfone, and Sasha Romanosky.
\newblock Common vulnerability scoring system.
\newblock {\em IEEE Security \& Privacy}, 4(6):85--89, 2006.

\bibitem{10.5555/2600239.2600241}
Dirk Merkel.
\newblock Docker: Lightweight linux containers for consistent development and
  deployment.
\newblock {\em Linux J.}, 2014(239), March 2014.

\bibitem{moore2012metasploitable}
HD~Moore.
\newblock Metasploitable 2 exploitability guide.
\newblock {\em Retrieved June}, 27:2013, 2012.

\bibitem{mp2016enhancing}
Amith~Raj MP, Ashok Kumar, Sahithya~J Pai, and Ashika Gopal.
\newblock Enhancing security of docker using linux hardening techniques.
\newblock In {\em 2016 2nd International Conference on Applied and Theoretical
  Computing and Communication Technology (iCATccT)}, pages 94--99. IEEE, 2016.

\bibitem{nilsson2006vulnerability}
Johan Nilsson and Vesa Virta.
\newblock Vulnerability scanners.
\newblock {\em Master of Science Thesis at Department of Computer and Systems
  Sciences, Royal Institute of Technology, Kista, Sweden}, 2006.

\bibitem{scarfone2008guide}
Karen Scarfone, Wayne Jansen, and Miles Tracy.
\newblock Guide to general server security.
\newblock {\em NIST Special Publication}, 800(s 123), 2008.

\bibitem{stern2012exploring}
Roni~Tzvi Stern, Meir Kalech, Alexander Feldman, and Gregory Provan.
\newblock Exploring the duality in conflict-directed model-based diagnosis.
\newblock In {\em Twenty-Sixth AAAI Conference on Artificial Intelligence},
  2012.

\bibitem{stumptner96aModelBased}
Markus Stumptner and Franz Wotawa.
\newblock A model-based approach to software debugging.
\newblock In {\em the Seventh International Workshop on Principles of Diagnosis
  (DX)}, pages 214--223, 1996.

\bibitem{sutton2018reinforcement}
Richard~S Sutton and Andrew~G Barto.
\newblock {\em Reinforcement learning: An introduction}.
\newblock MIT press, 2018.

\bibitem{DeepExploit}
Isao Takaesu.
\newblock Deepexploit, from
  https://github.com/13o-bbr-bbq/machine\textunderscore learning\textunderscore
  security/tree/master/deepexploit.

\bibitem{tassey2002economic}
Gregory Tassey.
\newblock The economic impacts of inadequate infrastructure for software
  testing.
\newblock {\em National Institute of Standards and Technology, RTI Project},
  7007(011):429--489, 2002.

\bibitem{team2003pax}
PaX Team.
\newblock Pax non-executable pages design \& implementation.
\newblock {\em Avaliable: http://pax. grsecurity. net}, 2003.

\bibitem{team2003address}
PaX Team et~al.
\newblock Address space layout randomization (aslr).
\newblock {\em ht tp://pax. grsecurity. net/docs/aslr. txt}, 2003.

\bibitem{turnbull2006hardening}
James Turnbull.
\newblock {\em Hardening Linux}.
\newblock Apress, 2006.

\bibitem{watkins1989learning}
Christopher John Cornish~Hellaby Watkins.
\newblock Learning from delayed rewards.
\newblock 1989.

\bibitem{wotawa2002creatingmodelformbd}
Franz Wotawa, Markus Stumptner, and Wolfgang Mayer.
\newblock Model-based debugging or how to diagnose programs automatically.
\newblock In {\em Proc. Int. Conf. Ind. Eng., Appl. Artif. Intell. Expert
  Syst.,}, page 746–757, 2002.

\end{thebibliography}

\onecolumn
\appendix
\section{\label{sec:appendix}Environmental conditions that may affect the exploitability of vulnerabilities.}
\begin{table*}[h]
\centering
\begin{tabular}{|p{0.13\textwidth}|p{0.35\textwidth}|p{0.4\textwidth}|}
\hline
{\textbf{Group}} & {\textbf{Environmental condition}} & {\textbf{Enabled environmental condition}} \\ \hline
\multirow{12}{*}{\textbf{Access Control}} & \multirow{4}{*}{User accounts} & Remove/Add guest users \\ \cline{3-3} 
 &  & Remove/Add users with empty passwords \\ \cline{3-3} 
 &  & Remove/Add users with naive passwords \\ \cline{3-3} 
 &  & Lock/Unlock users \\ \cline{2-3} 
 & Status of mandatory access control (MAC) & Enable/Disable SELinux Kernel \\ \cline{2-3} 
 & \multirow{3}{*}{File permissions} & Allow/Prevent users to read/write system files \\ \cline{3-3} 
 &  & Use noexec, nodev, nosuid on network services \\ \cline{3-3} 
 &  & Prevent users from writing to $/boot$ directory \\ \cline{2-3} 
 & \multirow{2}{*}{Root users} & Add/Delete users with UID=0 \\ \cline{3-3} 
 &  & Add/Remove users from sudoers group \\ \cline{2-3} 
 & Status of root login &  Modify the $/etc/passwd$ file\\ \cline{2-3} 
 & Separating user from OS & Move~$/proc$, $/bin$, $/sbin$ to a second disk partition \\ \hline
\multirow{4}{*}{\textbf{Connectivity}} & Open ports & Open/Close telnet and ftp ports \\ \cline{2-3} 
 & Status of encryption & Enable/Disable VPN or GnuPG \\ \cline{2-3} 
 & Status of ICMP or broadcast requests &  Modify the relevant kernel variable for blocking ICMP message\\ \hline
\multirow{3}{*}{\textbf{Services}} & Running services & Run/Stop Web services \\ \cline{2-3} 
 & Status of isolation mechanisms & Enable/Disable chroot \\ \hline
\multirow{3}{*}{\textbf{Safeguards}} & Status of memory protection
mechanisms &  Enable address space layout randomization (ASLR)  \\ \cline{2-3} 
 & Status of stack protectors & Enable Canary \\ \hline
 \multirow{2}{*}{\textbf{Packages}} & Installed packages & Delete flash player \\ \cline{2-3} 
 & Status of desktop environments &  Enable/Disable KDE/GNOME \\  \hline
 
\end{tabular}%
\end{table*}

\end{document}